\documentclass[twocolumn]{aastex}
\usepackage{graphics}
\usepackage{tabu}
\usepackage{natbib}
\usepackage{color}
\usepackage{rotating}
\usepackage{multirow}
\usepackage{footnote}
\usepackage{datetime}
\usepackage{amsmath}
\usepackage{mathrsfs}
\usepackage[normalem]{ulem}

\begin{document}

\title{Direct Detection of Black Hole-Driven Turbulence in the Centers of Galaxy Clusters} 

\author{Yuan Li}
\affil{Department of Astronomy, University of California, Berkeley, CA 94720, USA}
\email{yuan.astro@berkeley.edu}

\author{Marie-Lou Gendron-Marsolais}
\affil{European Southern Observatory, Alonso de C\'{o}rdova 3107, Vitacura, Casilla 19001, Santiago, Chile}

\author{Irina Zhuravleva}
\affil{Department of Astronomy \& Astrophysics, University of Chicago, 5640 S Ellis Ave, Chicago, IL 60637, USA}

\author{Siyao Xu}
\affil{Department of Astronomy, University of Wisconsin, 475 North Charter Street, Madison, WI 53706, USA}

\author{Aurora Simionescu}
\affil{SRON Netherlands Institute for Space Research Sorbonnelaan 2, 3584 CA Utrecht, The Netherlands}
\affil{Leiden Observatory, Leiden University, PO Box 9513, 2300 RA Leiden, The Netherlands}
\affil{Kavli Institute for the Physics and Mathematics of the Universe (WPI), University of Tokyo, Kashiwa 277-8583, Japan}

\author{Grant R. Tremblay}
\affil{Harvard-Smithsonian Center for Astrophysics, 60 Garden St., Cambridge, MA 02138, USA}

\author{Cassandra Lochhaas}
\affil{Department of Astronomy, The Ohio State University, 140 West 18th Avenue, Columbus, OH 43210, USA}
\affil{Space Telescope Science Institute, 3700 San Martin Dr., Baltimore, MD 21218, USA}

\author{Greg L. Bryan}
\affil{Department of Astronomy, Columbia University, 550 West 120th Street, New York, NY 10027, USA}
\affil{Center for Computational Astrophysics, Flatiron Institute, 162 5th Avenue, New York, NY 10010, USA}

\author{Eliot Quataert}
\affil{Department of Astronomy, University of California, Berkeley, CA 94720, USA}

\author{Norman W. Murray}
\affil{Canadian Institute for Theoretical Astrophysics, 60 St George Street, University of Toronto, ON M5S 3H8, Canada}
\affil{Canada Research Chair in Astrophysics}

\author{Alessandro Boselli}
\affil{Aix Marseille Univ, CNRS, CNES, LAM, Marseille, France}

\author{Julie Hlavacek-Larrondo}
\affil{D\'{e}partement de Physique, Université de Montr\'{e}al, Succ. Centre-Ville, Montr\'{e}al, Qu\'{e}bec, H3C 3J7, Canada}

\author{Yong Zheng}
\affil{Department of Astronomy, University of California, Berkeley, CA 94720, USA}

\author{Matteo Fossati}
\affil{Institute for Computational Cosmology and Centre for Extragalactic Astronomy, Department of Physics, Durham University, South Road, Durham DH1 3LE, UK}

\author{Miao Li}
\affil{Center for Computational Astrophysics, Flatiron Institute, 162 5th Avenue, New York, NY 10010, USA}

\author{Eric Emsellem}
\affil{European Southern Observatory, Karl-Schwarzschild-Strasse 2, 85748 Garching, Germany}
\affil{Univ Lyon, Univ Lyon1, ENS de Lyon, CNRS, Centre de Recherche Astrophysique de Lyon, UMR5574, F-69230 Saint-Genis-Laval France}

\author{Marc Sarzi}
\affil{Armagh Observatory and Planetarium, College Hill, Armagh BT61 9DG, Northern Ireland}

\author{Lev Arzamasskiy}
\affil{Department of Astrophysical Sciences, Princeton University, Ivy Lane, Princeton, NJ 08540}

\author{Ethan T. Vishniac}
\affil{Department of Physics \& Astronomy, Johns Hopkins University, Baltimore, MD, USA}

\begin{abstract}

Supermassive black holes (SMBHs) are thought to provide energy that prevents catastrophic cooling in the centers
of massive galaxies and galaxy clusters. However, it remains unclear how this ``feedback'' process operates. We use
high-resolution optical data to study the kinematics of multi-phase filamentary structures by measuring the velocity
structure function (VSF) of the filaments over a wide range of scales in the centers of three nearby galaxy clusters: Perseus, Abell 2597 and Virgo. 
We find that the motions of the filaments are turbulent in all three clusters studied. There is a clear correlation between features of the VSFs and the sizes of bubbles
inflated by SMBH driven jets. Our study demonstrates that SMBHs are the main driver of turbulent gas motions in the
centers of galaxy clusters and suggests that this turbulence is an important channel for coupling feedback to the
environment. 
Our measured amplitude of turbulence is in good agreement with Hitomi Doppler line broadening measurement and X-ray surface brightness fluctuation analysis, suggesting that the motion of the cold filaments is well-coupled to that of the hot gas. 
The smallest scales we probe are comparable to the mean free path in the intracluster medium (ICM). 
Our direct detection of turbulence on these scales provides the clearest evidence to date that isotropic viscosity is 
suppressed in the weakly-collisional, magnetized intracluster plasma.
\end{abstract}

\section{Introduction}
\label{sec:intro} 
\setcounter{footnote}{0} 

Relaxed galaxy clusters often harbor a cool core, where radiative cooling of the ICM is expected to result in cooling flows of hundreds of $\rm M_\odot \, yr^{-1}$ in the absence of heating \citep{Fabian1994}. Feedback from active galactic nuclei (AGN) in forms of jets, radiation, and fast outflows is thought to provide the energy to balance radiative cooling and suppress star formation \citep{McNamara2005}. X-ray observations show that AGN feedback generates ``bubbles'' and ``ripples'' in the surrounding intra-cluster medium (ICM) \citep{Fabian2012}. Based on X-ray measurements of line widths \citep{Hitomi2016} and surface brightness fluctuations \citep{Zhuravleva2014, Zhuravleva2016}, it is suggested that cluster cores are turbulent. However, current X-ray observatories have limited spatial and spectral resolutions, making it impossible to probe turbulence directly, let alone its drivers.

The centers of cool-core clusters also frequently exhibit extended filamentary structures that can be seen in the H$\alpha$ \citep{Conselice2001, Olivares2019} and sometimes CO \citep{Edge2001, McNamara2014}. The existence of cold filaments has been linked to the activities of SMBHs in the centers of galaxy clusters \citep{Cavagnolo2008,Tremblay2016}. The filaments often show perturbed kinematics and a lack of ordered motion on large scales \citep{Sarzi2006, Gendron-Marsolais2018, Olivares2019}. In other words, the motion of the filaments appears turbulent. 

In this work, we study the turbulent nature of multi-phase filaments by measuring their VSFs in three nearby galaxy clusters: Perseus, Abell 2597 and Virgo. We describe the data and data processing in Section~\ref{sec:data}. In Section~\ref{sec:results}, we connect the turbulent motions of the filaments to the activities of SMBHs, and compare our measurements with the X-ray analysis. In Section~\ref{sec:discussions}, we discuss the puzzling features of the VSFs, the uncertainties of the analysis, and the implications of our results, including constraints on microscopic physics of the ICM. We conclude this work in Section~\ref{sec:conclusions}.

\section{Data Processing}\label{sec:data}

\begin{table*}
\caption{Summary of Data}\label{table:data}
\centering
\vskip 0.05in 
\begin{tabular}{|c|c|c|l}
\cline{1-3}
                   & H$\alpha$ (resolution$^a$, seeing limit) & CO (resolution, beam size)    &  \\ \cline{1-3}
Perseus & CFHT (255 pc, $\sim 1$ kpc)      & N/A   &   \\ \cline{1-3}
Abell 2597         & MUSE (0.3 kpc, $\sim 1.5$ kpc)     & ALMA (0.2 kpc, $\sim 0.9$ kpc) &   \\ \cline{1-3}
Virgo        & MUSE$^b$ (16 pc, $\sim 80$ pc)     & ALMA$^c$ &    \\ \cline{1-3}
\end{tabular}
\\
\vskip 0.1in 
\raggedright{Note. a. This is the pixel size of the velocity maps shown in Figure~\ref{fig:all}. b. MUSE data only covers the central $\sim 4$ kpc of Virgo, and does not include the outer filaments. c. ALMA has observed only one molecular complex at a projected distance of 3 kpc from the center of Virgo \citep{Simionescu2018}. } 
\end{table*}

The Perseus H$\alpha$ filaments were observed using the optical imaging Fourier transform spectrometer SITELLE at the Canada France Hawaii Telescope (CFHT) \citep{Gendron-Marsolais2018}. SITELLE has a spatial resolution of $0.321''\times0.321''$, and a spectral resolution of $\rm R=1800$. The original Perseus data cube was binned up by a factor of 2 to increase the signal-to-noise ratio. The ionized filaments in Virgo and Abell 2597 was observed using the Multi Unit Spectroscopic Explorer (MUSE) with a spatial sampling of $0.2''$ and a spectral resolution of $\rm R=3000$ \citep{Sarzi2018, Boselli2019,Tremblay2018}. For Perseus and Virgo, the velocity in each pixel of the velocity map is obtained as the peak of a Gaussian profile fit to the $\rm H\alpha+NII$ complex, and for Abell 2597, only $\rm H\alpha$ is used in the fit. In Perseus, a small region in the center with a radius of $6''$ is excluded from the fitting due to contamination from the AGN \citep{Gendron-Marsolais2018}. The molecular gas in Abell 2597 and Virgo was observed using the Atacama Large Millimeter/submillimeter Array (ALMA) with a spatial resolution of $0.37''$ \citep{Simionescu2018, Tremblay2018}. See Table~\ref{table:data} for a summary of data.

\begin{figure*}
\centering
\includegraphics[width=0.29\linewidth]{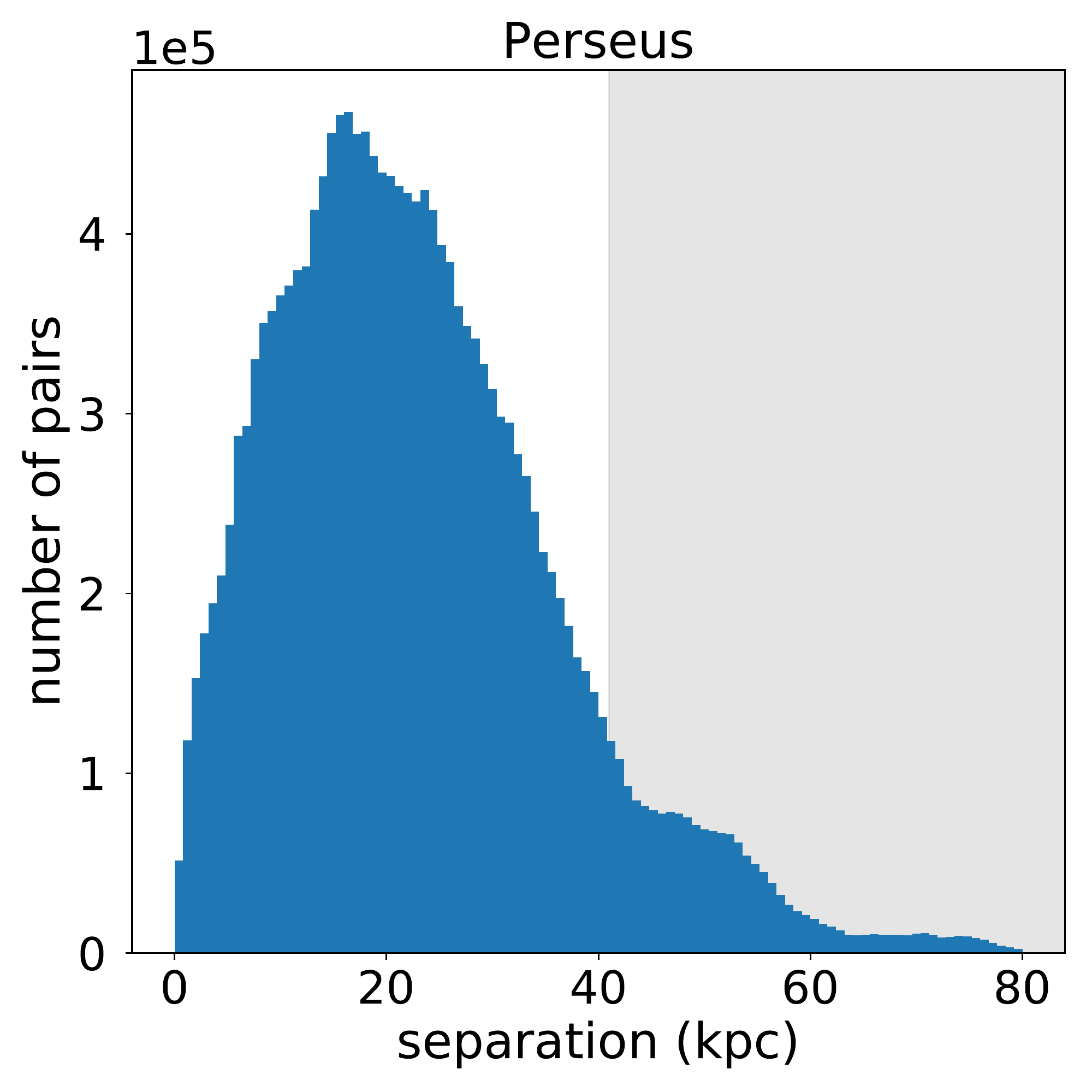}
\includegraphics[width=0.29\linewidth]{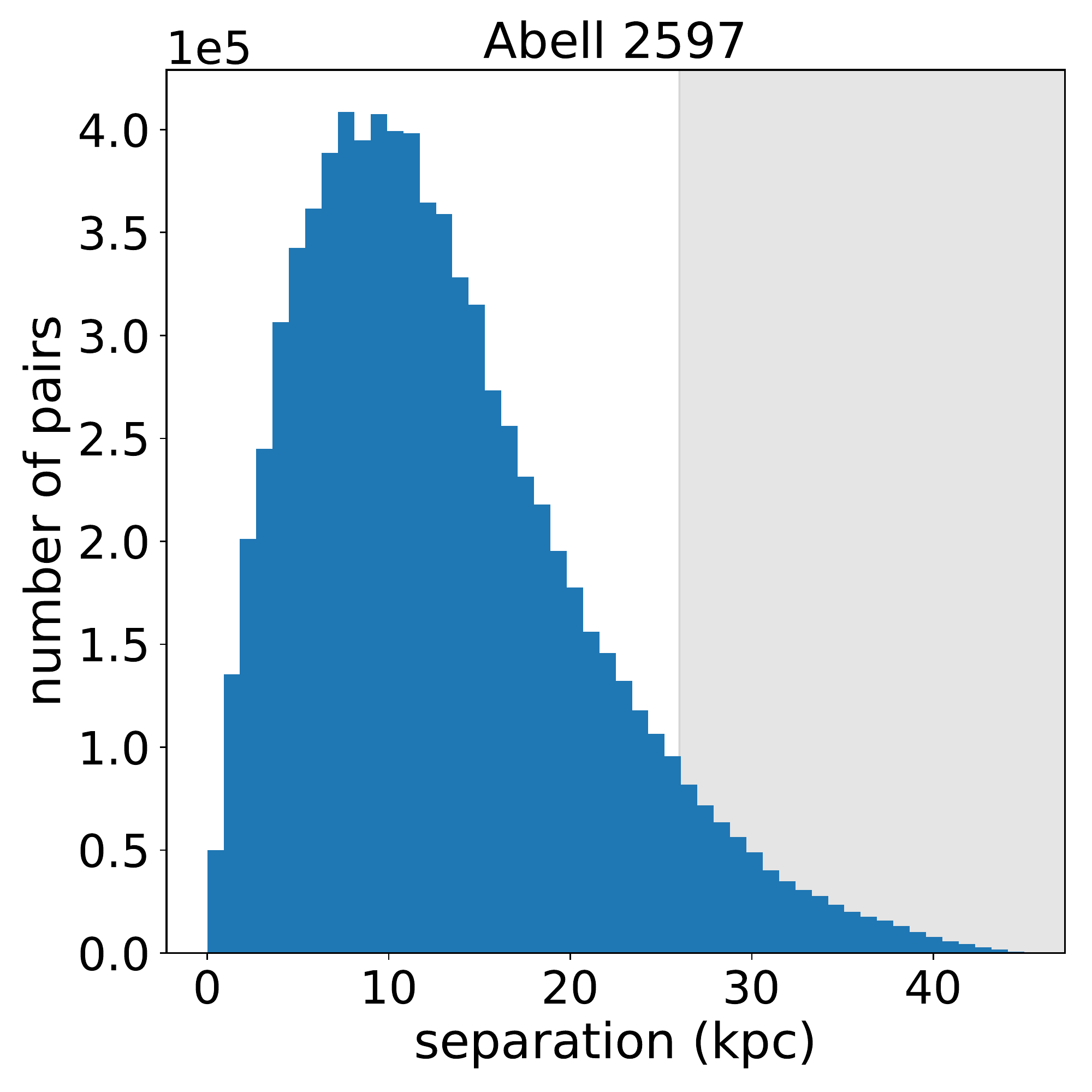}
\includegraphics[width=0.29\linewidth]{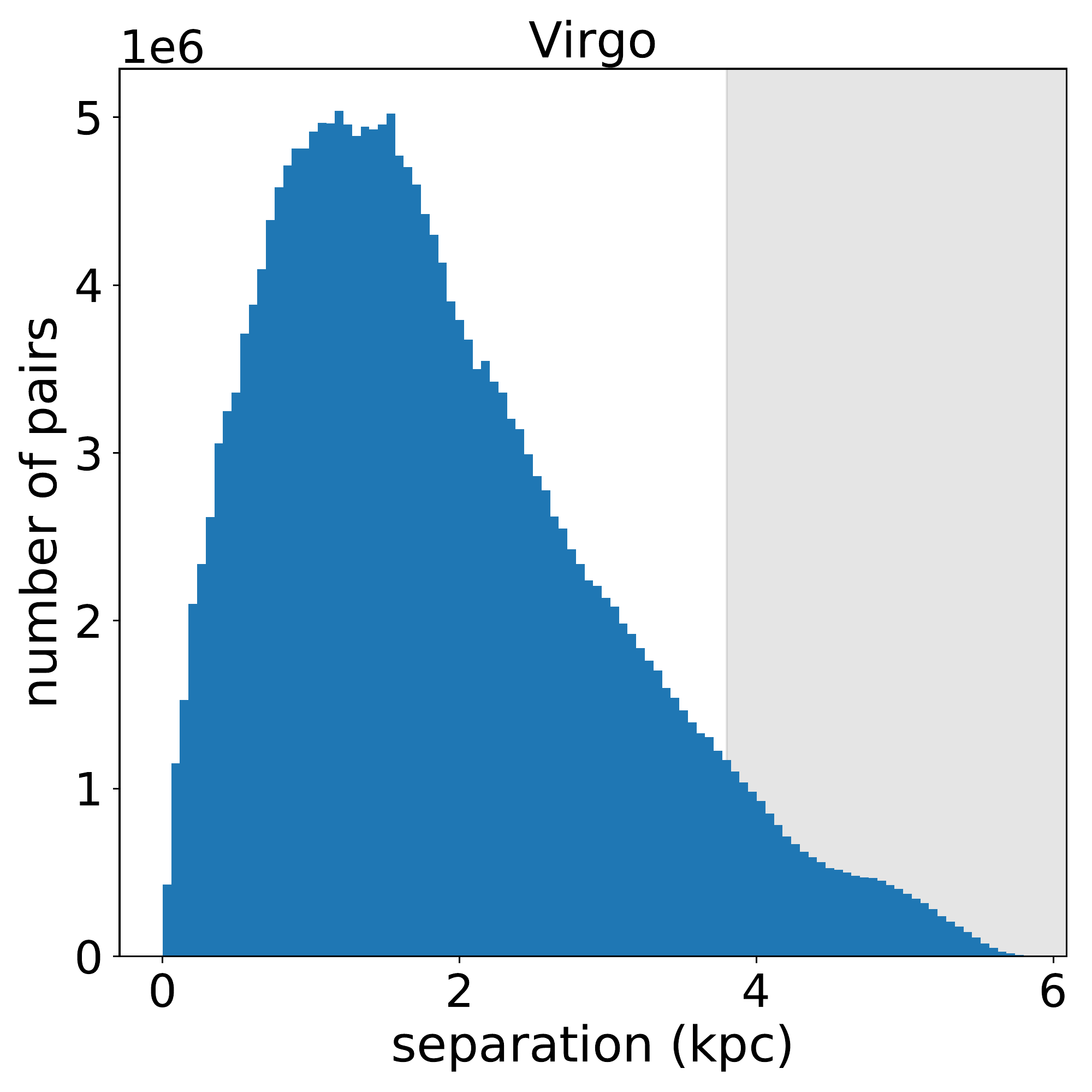}\\
\includegraphics[width=0.29\linewidth]{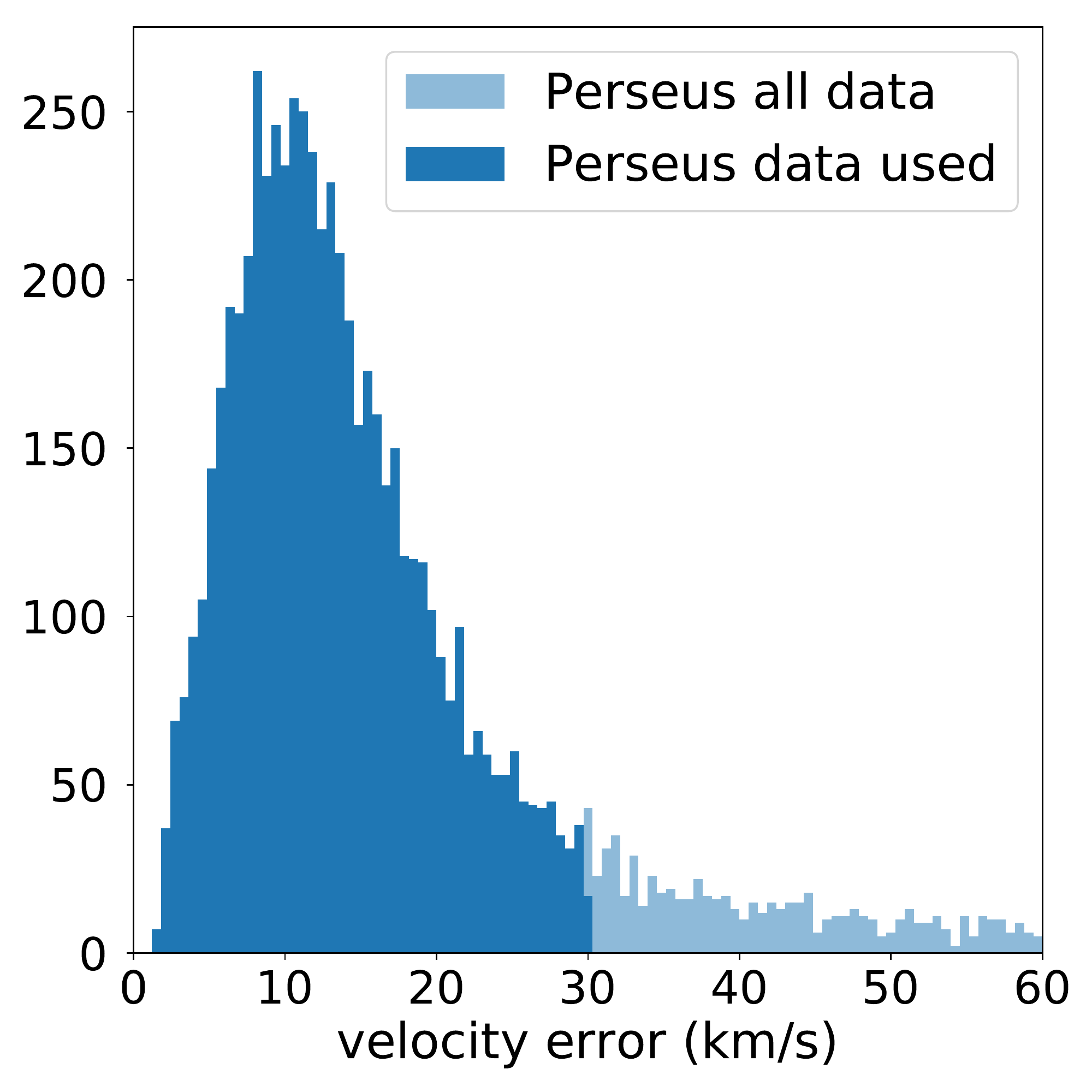}
\includegraphics[width=0.29\linewidth]{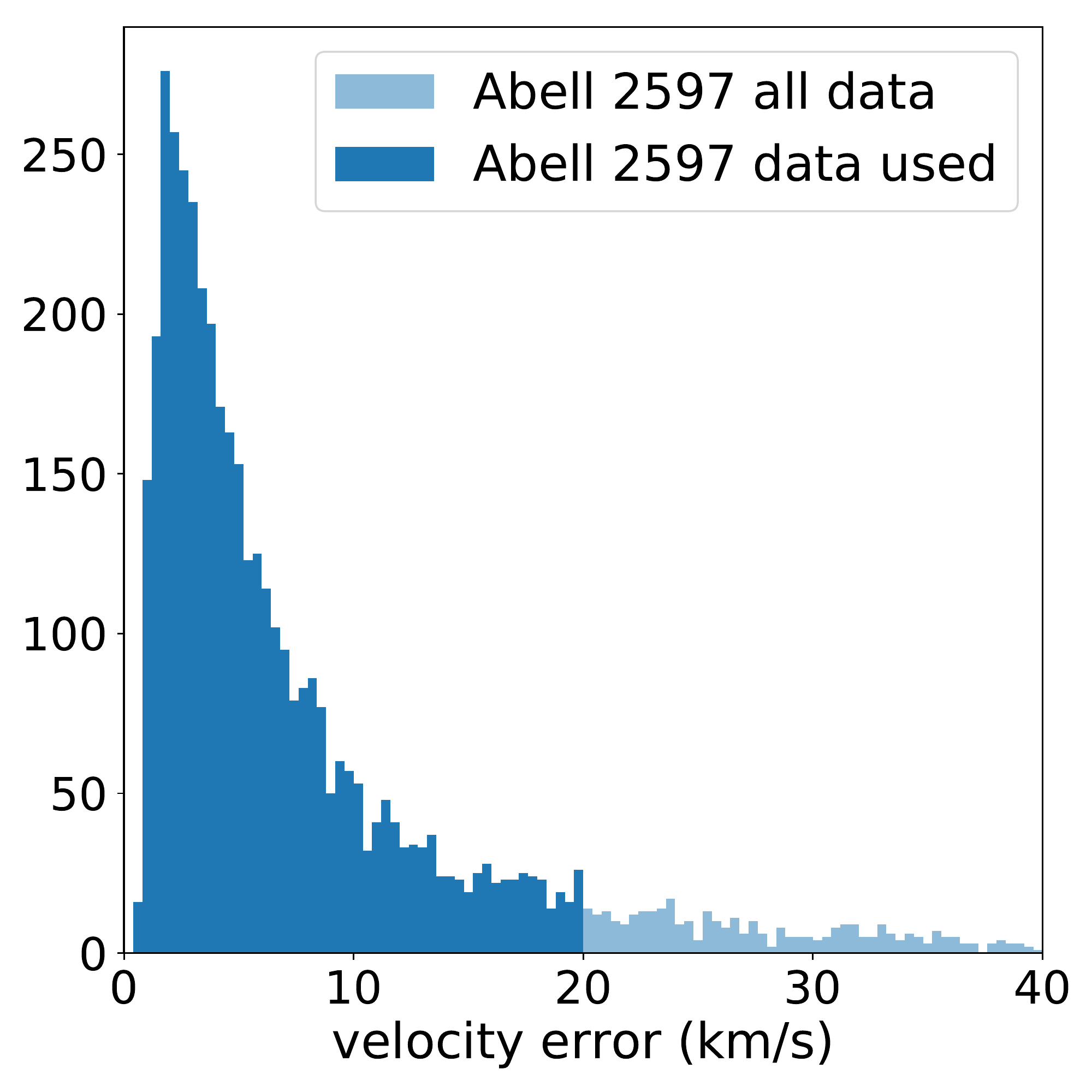}
\includegraphics[width=0.29\linewidth]{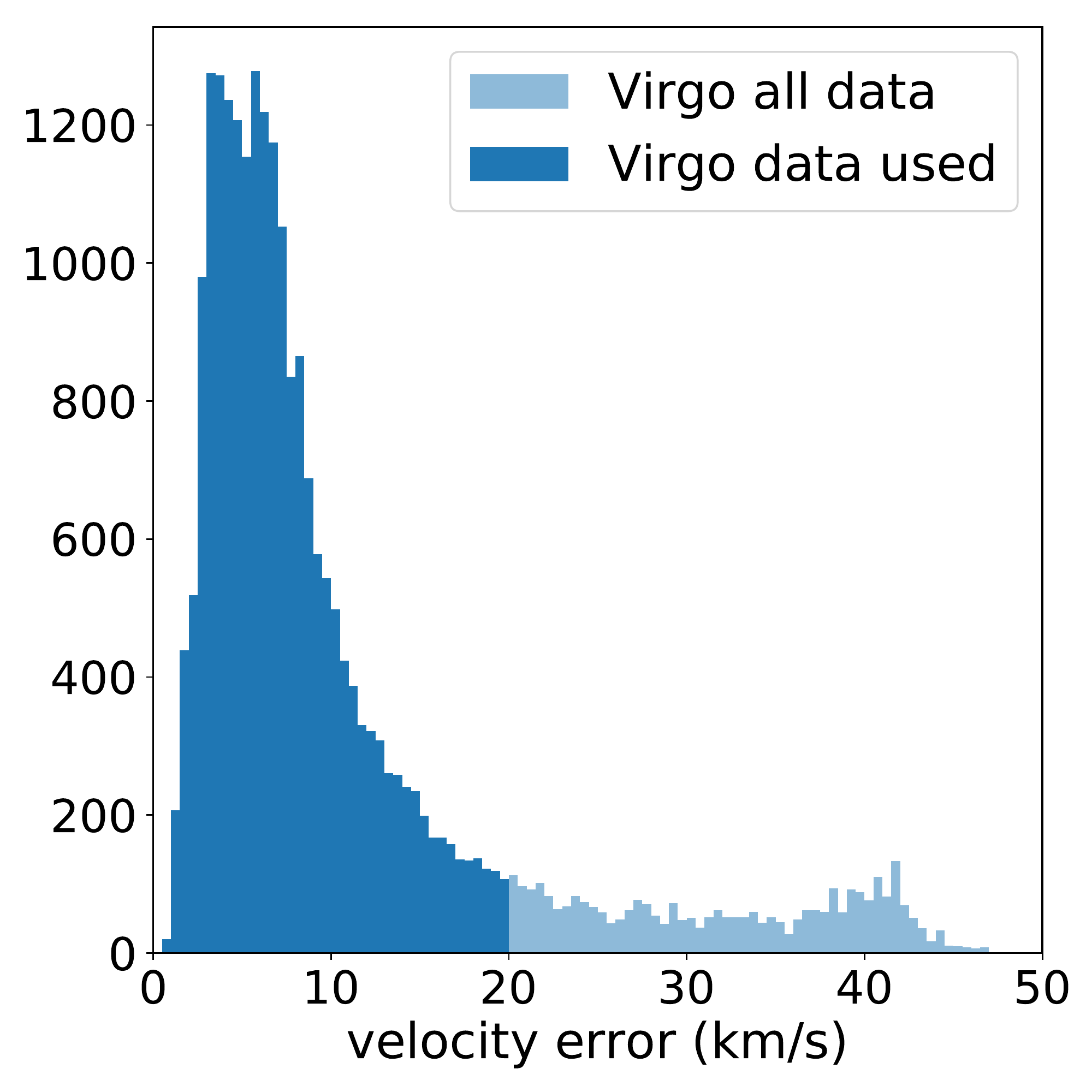}
\caption{Top left to right: distribution of velocity errors in Perseus, Abell 2597, and Virgo. Bottom left to right: distribution of pair separations. The grey areas denote where the number of pairs drops below 20\% of the peak. For Perseus and Abell 2597, the peak scales are $\sim 15$ kpc and $\sim 10$ kpc, respectively. They roughly correspond to the radius of the regions that contain most of the filaments. In Virgo, the peak scale corresponds to the size of the region observed by MUSE.}\label{fig:separation}
\end{figure*}

To understand the nature of the motion of these filaments, we compute the VSFs for all three clusters. We first remove a small fraction ($<20$\%) of pixels with large velocity errors, shown in the top panels of Figure~\ref{fig:separation}. We have visually examined pixels with very large velocity errors, and found that they tend to be located either at the edge of filaments or in isolation with an appearance similar to noise (even though it could be from a real gas cloud that is very faint and poorly resolved). Therefore, it is sensible to remove these pixels. The value of the velocity error cut is chosen to be a few times the median velocity error for each cluster. We have verified that the results are not sensitive to the exact choice of this value. For Perseus, an additional flux cut is applied to remove pixels with low signal-to-noise \citep{Gendron-Marsolais2018}. 

For each clean velocity map, we compute the first-order VSF in the following way: for each pair of pixels, we record the projected physical separation $\ell$ of the pair and compute the velocity difference $\delta v$ of the two pixels. The bottom panels of Figure~\ref{fig:separation} show the distribution of $\ell$. We then compute the average absolute value of the velocity differences $\langle |\delta v| \rangle$ within bins of $\ell$. The uncertainties in the VSFs are obtained by propagating the measurement errors. 

\section{Results}\label{sec:results}

\begin{figure*}
\centering
\includegraphics[scale=0.31,trim=0cm 0.3cm -2cm 0cm, clip=true]{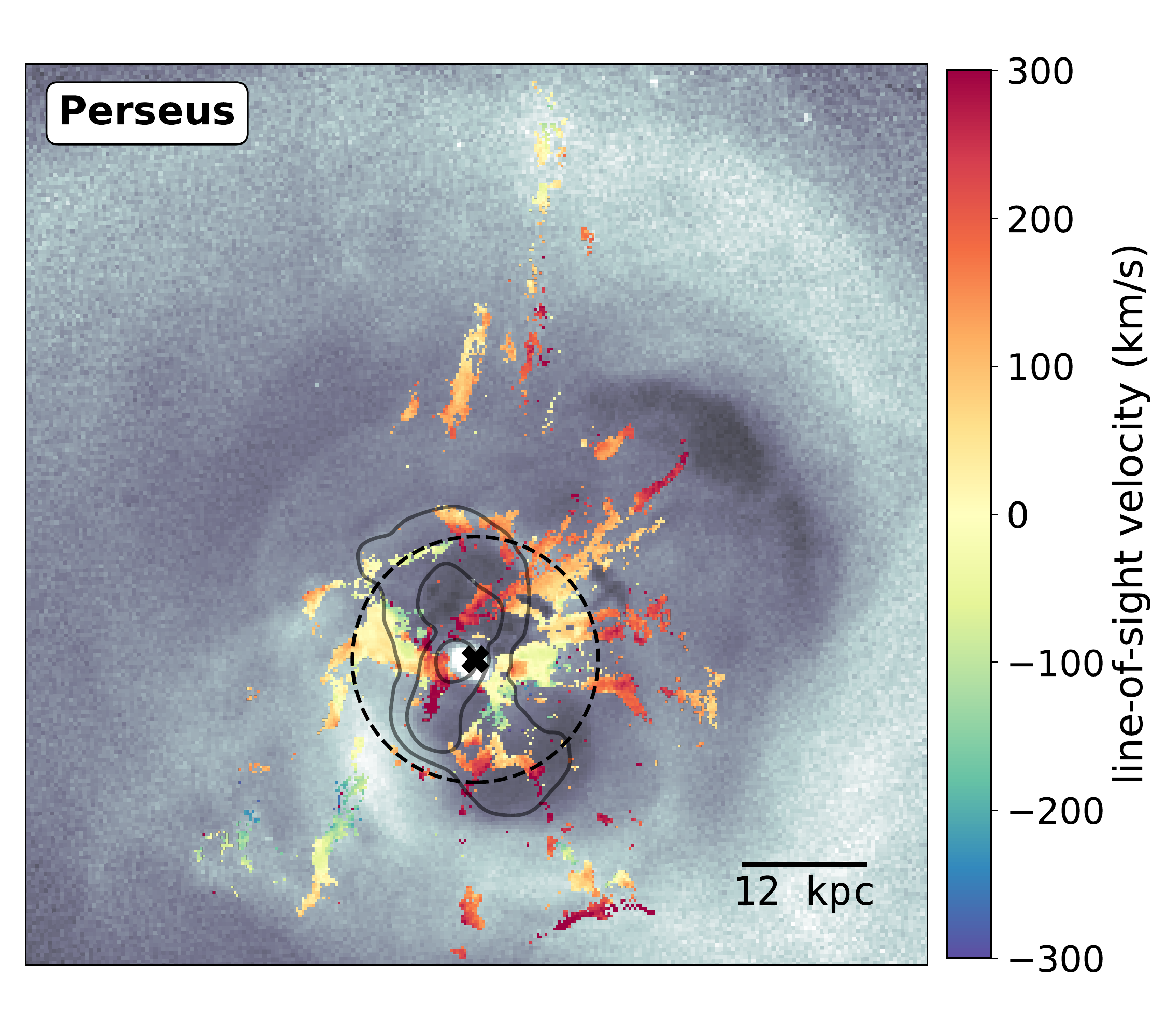}
\includegraphics[scale=0.298,trim=0cm 0.2cm 0cm 0cm, clip=true]{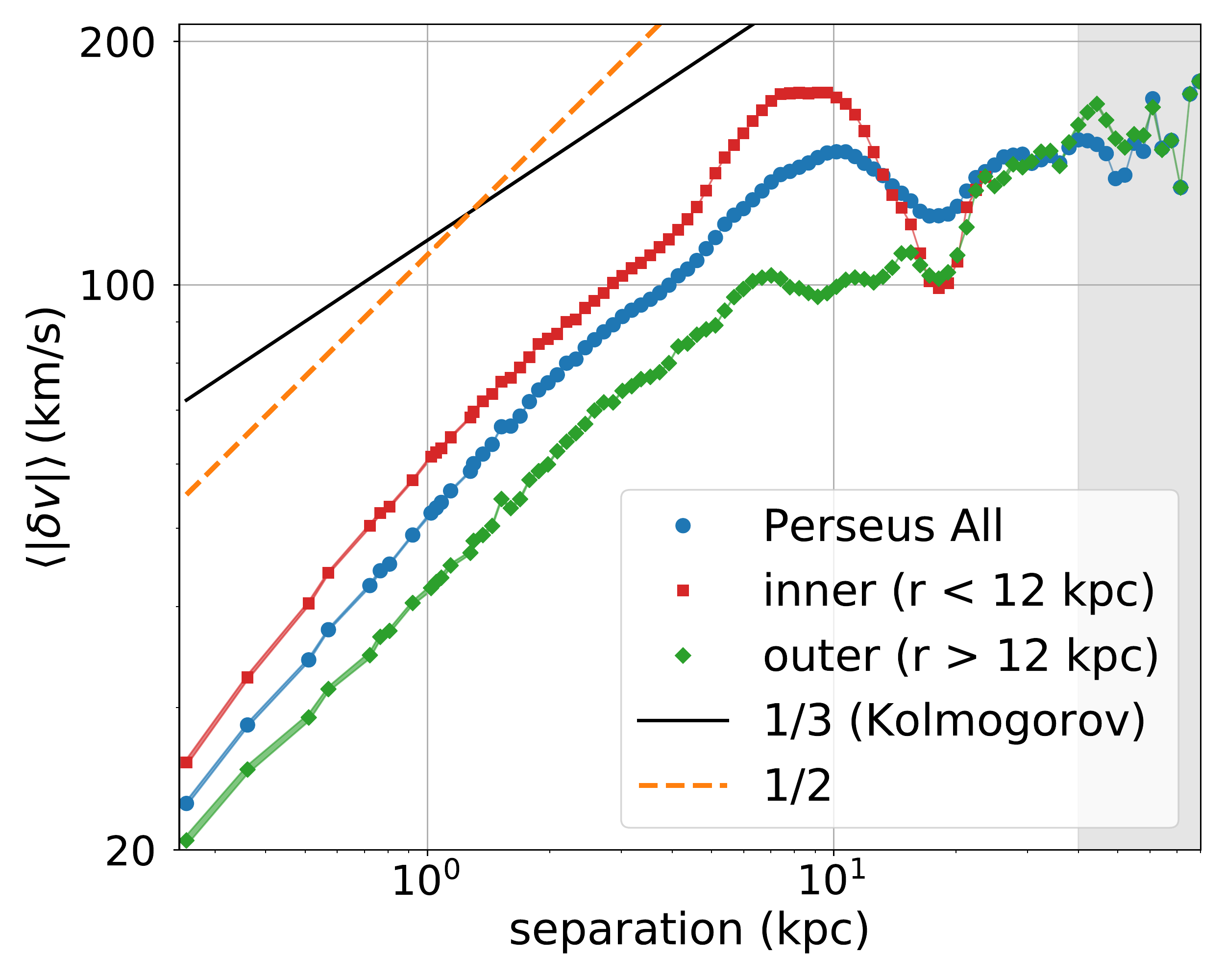}\\

\includegraphics[scale=0.31,trim=0cm 0.3cm -2cm 0.4cm, clip=true]{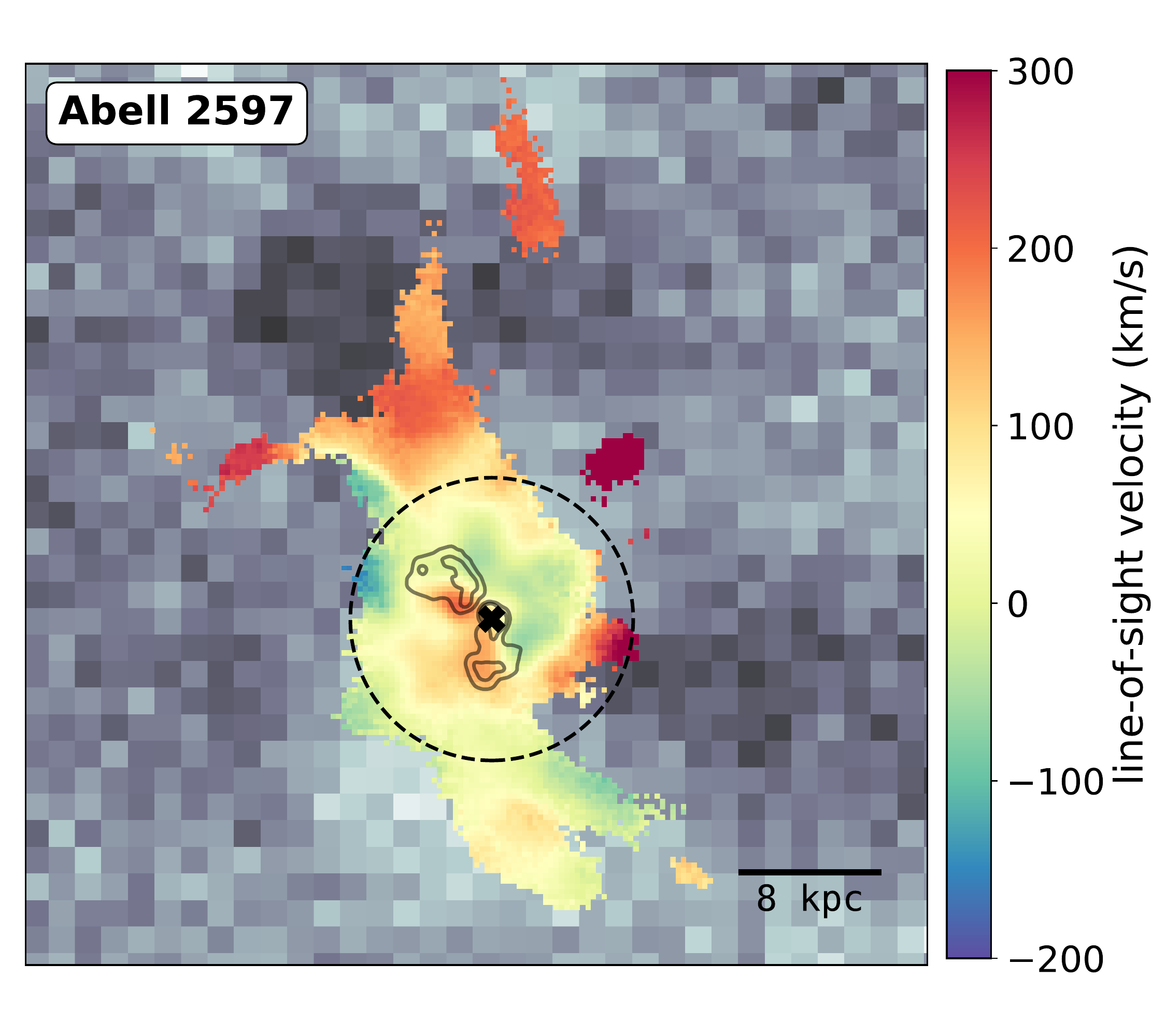}
\includegraphics[scale=0.298,trim=0cm 0.2cm 0cm 0.2cm, clip=true]{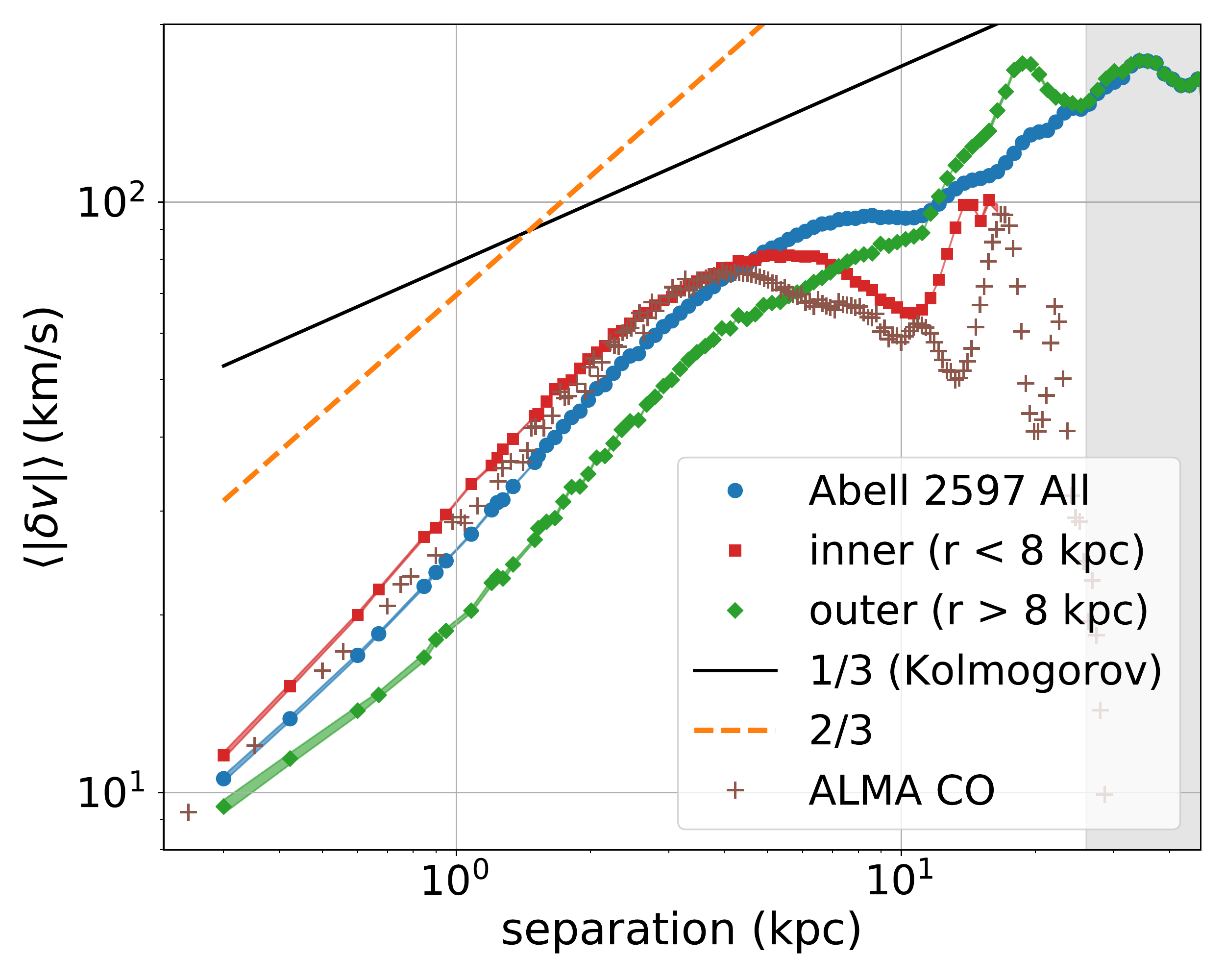}

\includegraphics[scale=0.31,trim=0cm 0cm -2cm 0.4cm, clip=true]{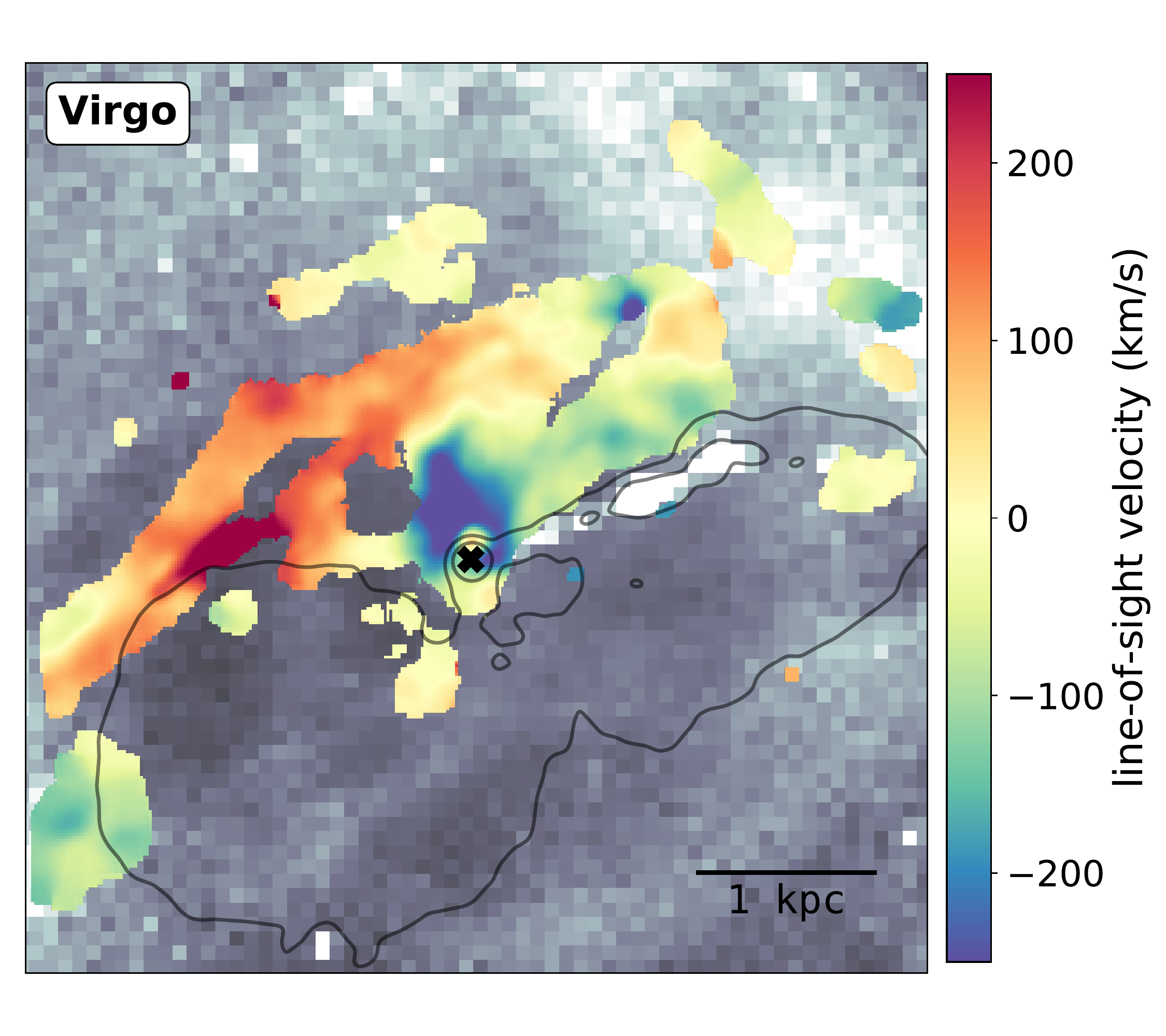}
\includegraphics[scale=0.298,trim=0cm 0.2cm 0cm 0.2cm, clip=true]{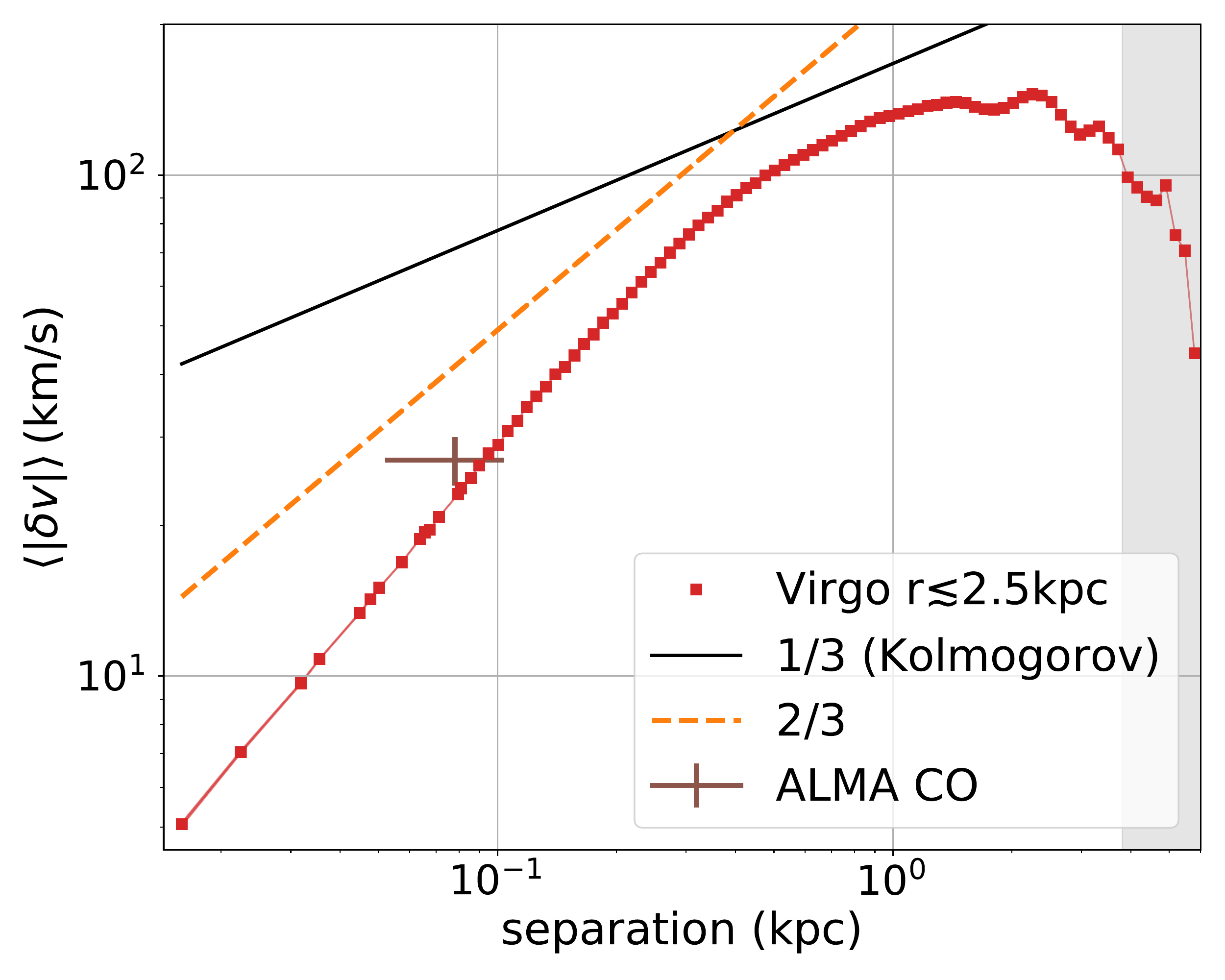}
\caption{Left: Velocity maps of the H$\alpha$ filaments overlaid on the X-ray residual images (shown in grey) in the centers of Perseus \citep{Gendron-Marsolais2018}, Abell 2597 \citep{Tremblay2018} and Virgo \citep{Sarzi2018, Boselli2019}. Right: Corresponding velocity structure functions (VSFs) of the filaments. In the left panels, the black cross indicates the position of the SMBH, and the black circles in Perseus and Abell 2597 denote the separation of the inner and outer regions in our analysis. Black contours show the low-frequency radio synchrotron emission. In the right panels, the thickness of the lines reflects the uncertainties from measurement errors. The grey areas denote where uncertainties from sampling limit are large. To guide the eye, we also plot solid black lines with a slope of 1/3 for Kolmogorov turbulence and orange dashed lines with steeper slopes. In all three clusters, the motion of the filaments is turbulent, and the features in the VSFs correspond to AGN activities. The VSF of Perseus reveals a driving scale of $\lesssim 10$ kpc, roughly the size of the inner bubbles. The two bumps in the VSF of Abell 2597 correspond to the inner and outer X-ray bubbles at $\sim 4$ and $\sim 20-30$ kpc. For Virgo, the inferred driving scale is $\sim 1-2$ kpc, roughly the size of the inner bubble and also the jet (the bright linear feature in the X-ray). The VSFs of the $\rm H\alpha$ filaments are consistent with those of the molecular gas observed by ALMA.
}\label{fig:all}
\end{figure*}

The left panels of Figure~\ref{fig:all} show the velocity maps of the H$\alpha$ filaments in Perseus, Abell 2597, and Virgo. The right panels show the corresponding VSFs. A broad power-law slope confirms the visual impression that the gas motion is turbulent. The VSFs of all three clusters show a flattening on scales above a characteristic pair separation, ranging from $\sim 10$ kpc (for Perseus) to $\sim 1-2$ kpc (for Virgo). The flattening of VSF indicates that this is the dominant driving scale of turbulence. Right below this characteristic scale, the slope of the VSF is $\sim 1/3$, and is consistent with the expectation of classical Kolmogorov turbulence for an incompressible fluid. On smaller scales, the slopes appear to be steeper, and vary from cluster to cluster (see Section~\ref{sec:steepening} for more discussions). 

To better reveal the driving source of turbulence, we divide the filaments in Perseus into inner ($r<12$ kpc) and outer filaments ($r>12$ kpc). We choose this dividing radius $r=12$ kpc such that there are comparable total numbers of pixels in the inner and the outer regions. We have verified that the results are not sensitive to the exact value of this radius. 

As the top right panel of Figure~\ref{fig:all} shows, the VSF of the inner filaments shows a similar shape as the VSF of all the filaments, but a larger amplitude and a more prominent break at $r\lesssim 10$ kpc. This is roughly the size of the inner X-ray bubbles of Perseus \citep{Fabian2003}, suggesting that the driver of turbulence is AGN feedback. On the other hand, the VSF of the outer filaments does not show a clear break at such a scale. Instead, the power continues to rise towards larger scales. This suggests that the outer filaments likely probe turbulence driven on larger scales. The VSF shows a bump at $20-30$ kpc, which is roughly the size of the outer bubble \citep{Fabian2003}. Thus it is possible that the turbulent motion of the outer filaments in Perseus is mainly caused by previous AGN outbursts. However, with current measurements, we cannot rule out the possibility that this area is dominated by turbulence driven by large-scale structure formation \citep{Ryu2008, ZuHone2018}.

\begin{figure}
\centering
\includegraphics[scale=0.33,trim=0cm 0cm 0cm 0cm, clip=true]{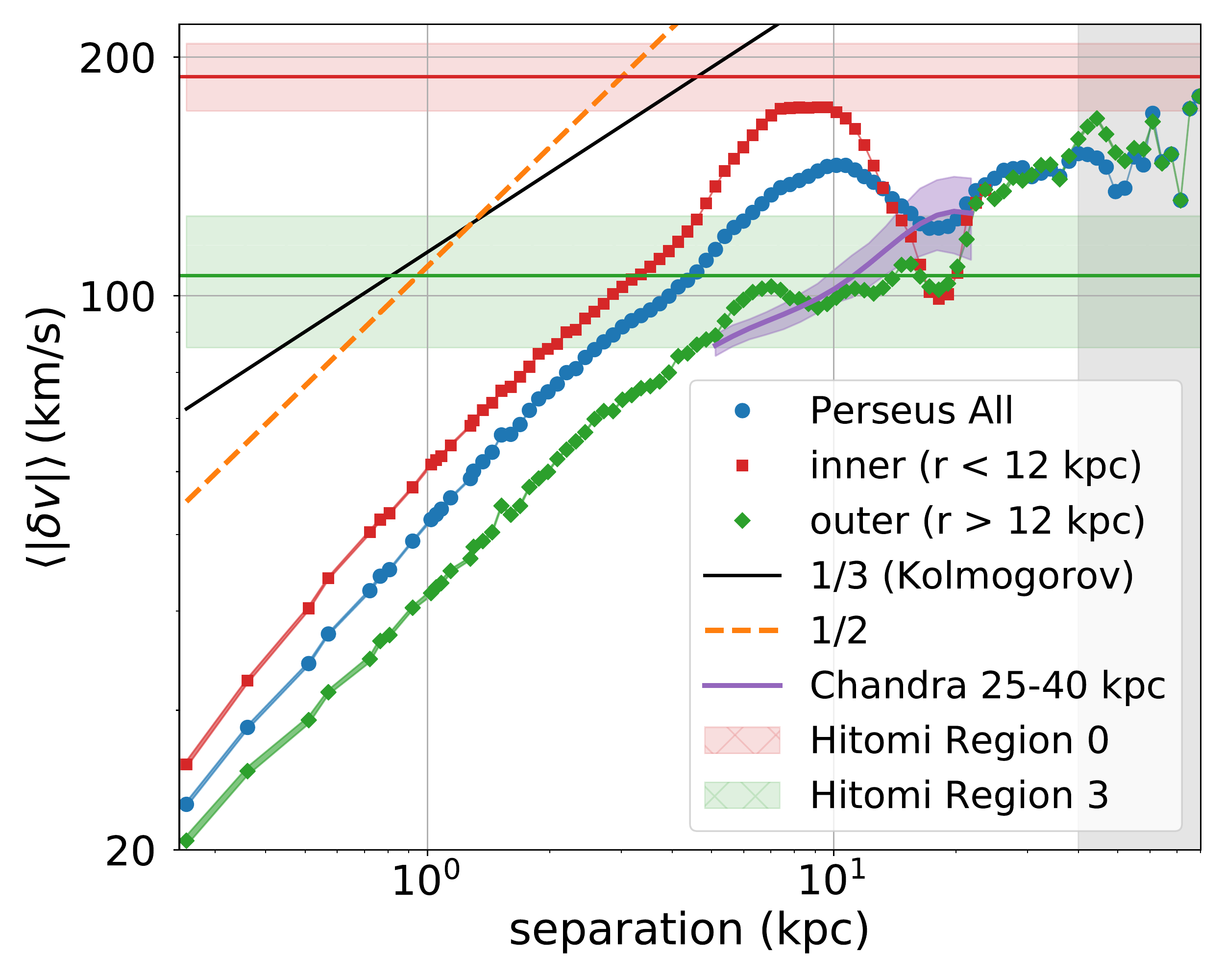}
\caption{Comparison with X-ray measurements of the Perseus cluster, including Hitomi \citep{Hitomi2016} X-ray Doppler line broadening measurements \citep{Hitomi2018} and Chandra surface brightness fluctuation analysis \citep{Zhuravleva2014}. The thickness of the lines reflects the uncertainties from measurement errors. Because Hitomi measures line broadening along the entire line-of-sight over a rather large projected area ($\sim 20$ kpc), we cannot derive a VSF from the measurements. Thus we show the Hitomi PSF corrected line-of-sight velocity dispersion measurements as horizontal lines with shaded regions reflecting the measurement uncertainties. Hitomi Region 0 roughly corresponds to our inner ($r<12$ kpc) region, and Hitomi Region 3 covers a large fraction of the outer filaments (corresponding to our $r>12$ kpc region) \citep{Hitomi2018}. The X-ray surface brightness fluctuation analysis for the $r<40$ kpc region excludes $r\lesssim 25$ kpc region due to presence of bubbles and shocks (see Appendix for more detailed discussions). Thus it roughly corresponds to the outer region of the $\rm H\alpha$ filaments. Our measured amplitudes of turbulence based on the optical data are in remarkable agreement with the X-ray results.}\label{fig:Hitomi}
\end{figure}

Hitomi has measured the line-of-sight velocity dispersion in the core of Perseus at much lower spatial resolution \citep{Hitomi2016}. Our measured velocities at and above the driving scale for the inner and outer filaments agree with the Hitomi measurements of the inner and outer regions \citep{Hitomi2018} of the Perseus core (Figure~\ref{fig:Hitomi}). 
In addition, the VSF of the outer filaments shows remarkable agreement with that inferred from the analysis of X-ray surface brightness fluctuations of similar regions \citep{Zhuravleva2014} (see Appendix for detail).

The inner filaments of Abell 2597 reveal a driving scale of $\sim 4$ kpc (middle panels of Figure~\ref{fig:all}), which is also seen in the VSF of the molecular gas observed by ALMA. The driving scale is again consistent with the size of the inner X-ray bubbles filled with radio-emitting plasma \citep{Tremblay2012}. For the outer filaments of Abell 2597, the power continues to rise towards larger separations. There is a clear bump between $20-30$ kpc, which is roughly the distance to the outer X-ray bubbles that are visible on the X-ray map. This feature is also seen in the VSF of the molecular gas. The X-ray observations of Abell 2597 show many shocks, bubbles, and ripples \citep{Tremblay2012}. It is likely that AGN-driven turbulence dominates the entire central region of Abell 2597. 

In Virgo (bottom panels of Figure~\ref{fig:all}), we again see a clear connection between AGN feedback and turbulence. The inferred driving scale in the center of Virgo is between 1-2 kpc, which is the size of the bright AGN jet \citep{Marshall2002} (the linear X-ray feature extending to the right) and also the jet-driven bubble. ALMA has observed a molecular complex located around the lower left corner of the map \citep{Simionescu2018}, and the measured velocity dispersion is in good agreement with our results. 

For all three clusters, the inferred driving scale is consistent with the scenario that AGN feedback is the main driver of turbulence in the centers of galaxy clusters. In addition, the amplitude of the turbulent motion revealed by the VSF is also consistent with this scenario. The largest velocity caused by AGN feedback is roughly the velocity of the post-shock material, which is $\frac{3}{2}(M-1)c_{\rm s}$ with $M$ being the Mach number of the shock and $c_{\rm s}$ being the sound speed of the ICM \citep{Li2017}. The measured $M$ in these clusters is $\sim 1.1-1.2$ \citep{Tremblay2012, Forman2017}, and $c_{\rm s}$ is a few hundred km/s. Therefore, the post-shock velocities are $\sim 100-200$ km/s. If turbulence is driven by buoyantly rising bubbles, the largest velocities should be the velocities of the bubbles, which are also expected to be a fraction of the sounds speed \citep{Robinson2004}.

\section{Discussions}\label{sec:discussions}
\subsection{The steepening of the VSF}\label{sec:steepening}
The steepening of the VSF on small scales is puzzling. For all three clusters, the steepening happens on scales well above the seeing limit (see Table~\ref{table:data} for a summary of the seeing limit), so it is a real feature (see Section~\ref{sec:uncertainties} for more detailed discussions on the effects of seeing and other uncertainties). A transition from subsonic turbulence to supersonic turbulence would steepen the slope from 1/3 to 1/2, as is seen in Perseus, and we do expect this transition to happen at some point within the cold filaments where the sound speed is low. However, the steepening happens on scales much larger than the typical width of the filaments ($< 1$ kpc) in Perseus\citep{Conselice2001}. Moreover, only Perseus has a $\sim 1/2$ slope on small scales, whereas the other two clusters show even steeper slopes, which cannot be explained by supersonic turbulence. 

We do not yet have a definitive explanation for the steepening and the exact slopes of the VSFs. There are, however, some interesting theoretical possibilities. On small scales (from near and below the mean free paths down to Larmor radii), gas motion is likely dominated by Alfv\'{e}n waves. It is possible that the steepening of the VSF is a result of partial dissipation of certain modes. Magnetic fields can also steepen the kinetic power spectrum if magnetic tension suppresses the nonlinear decay of g-modes \citep{Bambic2018}.  

Another interesting possibility is that the turbulence cascade is affected by kinetic micro-instabilities, such as firehose and mirror instabilities \citep{Kunz2014, Squire2019}. MHD waves, in particular, Alfv\'{e}n waves may become unstable to these instabilities \citep{Squire2017}. Turbulent energy in this case would be transferred non-locally from large scales to the much smaller lengths relevant for individual protons, which may result in a steeper spectrum. Future theoretical investigations are required to help understand how these instabilities affect the spectrum of turbulence.

It is also possible that we are seeing features unique to turbulence driven by intermittent AGN feedback. The eddy turnover time associated with scale $\ell$ can be estimated as $t_\ell \sim \ell/v_\ell$. Our analysis of Perseus reveals a driving scale $L\sim 10$ kpc, and the velocity at the driving scale is $v_L \sim 140$ km/s. Thus $t_L \sim 70$ Myr. The period of AGN outbursts can be estimated from the inferred age separation of X-ray bubbles \citep{Sanders2007}, which gives a period of $\sim 10$ Myr, much shorter than $t_L$. It takes a few $t_L$ for turbulence to cascade down from the driving scale $L$ to the dissipation scale, which means that the time it takes to establish a classic Kolmogorov turbulence is an order of magnitude longer than the intermittency of the driver. The same is true for Abell 2597 and Virgo. 

AGN feedback as a turbulence driver is not only intermittent (in the sense that it turns on and off on short time-scales compared with $t_L$), its strength, driving scale and the volume it influences also all change over time. Each outburst grows from small scales to large scales, as does its ``sphere of influence''. In this picture, the VSF steepening reflects a suppression of power on small scales, and can be explained by the fact that a fraction of the gas is not as perturbed. The less perturbed gas may have a Kolmogorov spectrum from the cascade of turbulence driven by structure formation, supernova type Ia, and previous AGN activities, but the amplitude is too low to be detected on scales we are able to probe with confidence here\footnote{There is a hint of flattening of the VSF on small scales in Abell 2597, especially for the outer filaments, which may be probing turbulence driven by structure formation.}. 

\subsection{Limitations and Uncertainties}\label{sec:uncertainties}
On small scales, optical observations are affected by ``seeing'' due to turbulence in the Earth atmosphere. Seeing may have a larger effect on the flux measurement, but less on the line-of-sight velocity measurement. The reason is that even though neighboring pixels would share photons due to seeing, the velocity measurement is only sensitive to the shift of the brightest component along the line of sight. If our velocity measurements were strongly affected by seeing, then one would expect a further steepening of the VSF on scales below the seeing limit. This is not observed in our results as the slope remains the same at the smallest scale measured. 

Another source of uncertainties has to do with overlapping filaments along the line-of-sight. In the central regions, an individual line-of-sight can probe multiple H$\alpha$ emitting clouds. For all the pixels, we always fit with one Gaussian component. We have individually inspected a large number of pixel fits in Virgo, and verified that in case there are two components along the line-of-sight (which are rare), the fit correctly locks onto the strongest component. Thus even though the velocity dispersion may become large due to overlapping filaments \citep{Gendron-Marsolais2018}, the centroid velocity probes only the velocity shift of the brightest filament, and is therefore robust. We also know that the outer filaments do not tend to have this overlapping issue \citep{Conselice2001}. The inner and outer filaments show similar VSFs for both Perseus and Abell 2597. This confirms that the overlapping issue does not significantly affect our analysis. 

However, we do think that our results can be affected by projection effect. That is, two pixels close to each other in projection may not be physically close to each other, and may show a rather large velocity difference. This affects the VSF on smaller scales more, due to smaller number of pairs and smaller intrinsic velocity differences. Removing the projection effect requires an understanding of the true three-dimensional distribution of the filaments, which we currently do not have. The corrected slope would likely be even steeper than what we show here, but would not change our main conclusions. 

On large scales, our measurements suffer from the sampling limit. As Figure~\ref{fig:separation} shows, the total number of pairs decreases as the separation gets larger than the size of the whole H$\alpha$ structure. Thus at large separations, we are only sampling a small fraction of the whole volume, which can cause a bias. The grey areas in 
the top panels of Figure~\ref{fig:separation} denote where the number of pairs drops below 20\% of the peak, and the sampling uncertainties are considered large. They correspond to the grey areas in Figure~\ref{fig:all}. 
To better assess the uncertainties associated with the sampling limit, we have also examined the distribution of $\delta v$ at different scales. On scales where we consider sampling uncertainties to be large, the absolute value of the skewness tends to increase above $\sim 0.5-1$. Therefore, we caution against over-interpretation of features in the VSFs on very large scales. 

Overall, our results do not appear to be significantly affected by the limitations and uncertainties discussed here. Future optical observations with even better spatial and spectral resolutions will help improve the assessment of these uncertainties.

\subsection{Implications}

\begin{figure}
\centering
\includegraphics[scale=0.33,trim=0cm 0cm 0cm 0cm, clip=true]{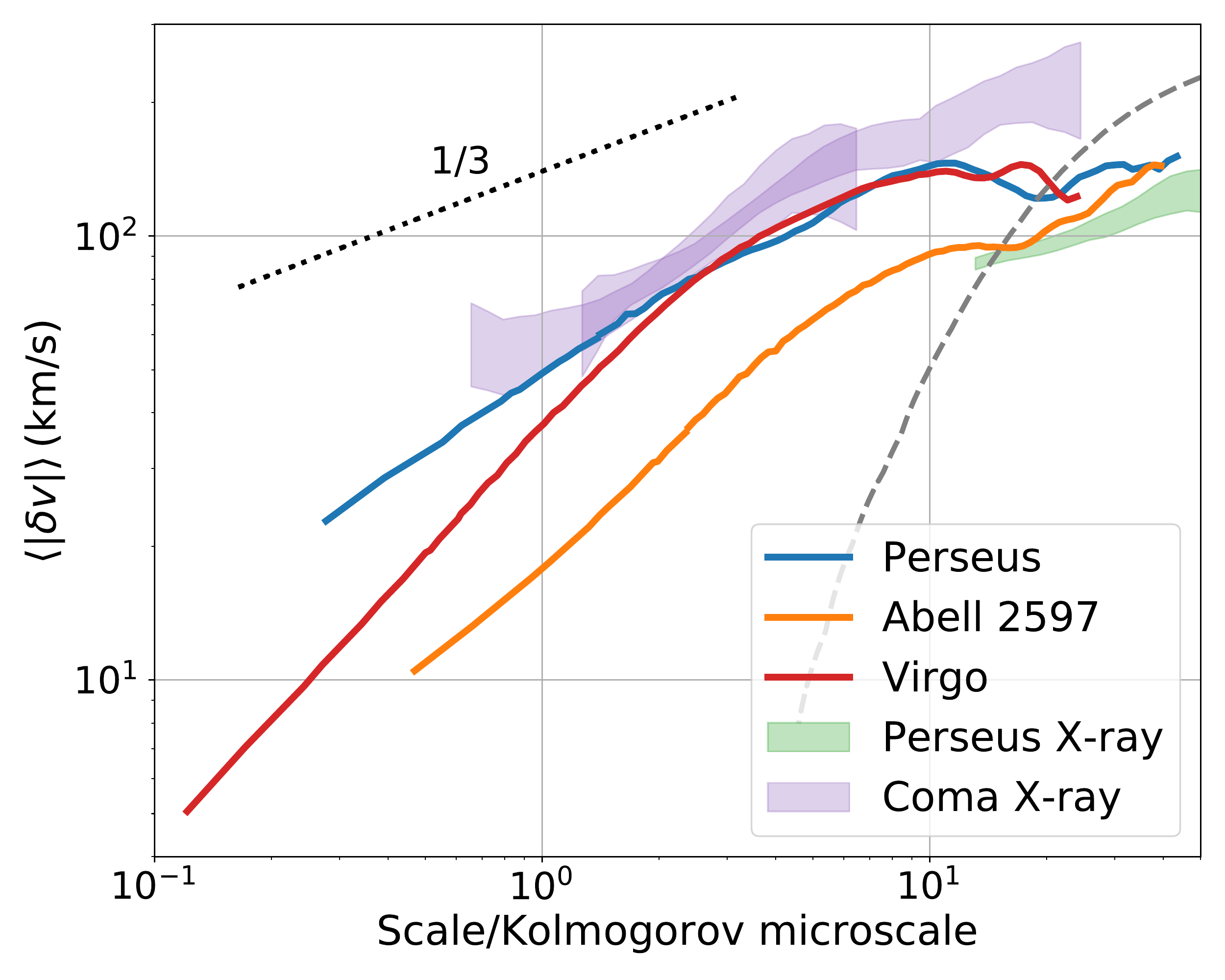}
\caption{VSFs with scales normalized by the Kolmogorov microscales. Also shown are the best constraints obtained previously using the X-ray surface brightness fluctuation analysis of the Coma cluster. For comparison, we have also plotted the Perseus X-ray analysis for the $r<40$ kpc region (excluding $r\lesssim 25$ kpc). The width of the X-ray curves shows 1$\sigma$ statistical uncertainties. The dashed grey line shows the prediction from direct numerical simulations (DNS) of hydrodynamic turbulence with Spitzer viscosity\citep{Ishihara2016}. Our direct detection of turbulence below the Kolmogorov microscales confirms the previous interpretation of the X-ray surface brightness analysis: the effective viscosity in the ICM is suppressed.}\label{fig:Irina}
\end{figure}

Our results suggest that the motion of cold filaments is well-coupled with the hot ICM. The origin of the $\rm H\alpha$ filaments and their fate are still uncertain, but two scenarios would allow the filaments to share the same turbulent motion of the hot ICM: (1) if they originate from the hot gas, either due to thermal instabilities or induced cooling \citep{McCourt2012, PII, Li2019}, but are very short-lived (dissolve quickly) such that they keep the memory of the turbulent motion of the hot gas, and/or (2) if they are very ``misty'' and quickly become co-moving with the hot gas \citep{McCourt2018} even if they are created independently of it \citep{Qiu2019}. On the other hand, if the cold gas is poorly coupled to the hot gas and follows ballistic trajectories, neighboring cold filaments would move independently and show little kinematic correlation. The measured VSF on small scales would be flatter than Kolmogorov, and certainly flatter than what is measured here. 

In addition, we can use the turbulent motion of the cold gas to put constraints on microscopic transport processes in the hot ICM. Figure~\ref{fig:Irina} shows velocities as a function of scales normalized by the Kolmogorov microscales. The Kolmogorov microscale where the turbulent kinetic energy is dissipated into heat, and is calculated as $\eta=\left (\frac{\nu^3}{\epsilon}\right)^{1/4}$, where $\nu$ is the kinematic viscosity and $\epsilon$ is the energy dissipation rate. The dynamic viscosity $\mu$, which is related to the kinetic viscosity as $\mu=\rho \nu$, can be estimated as:
\begin{equation}
\mu=5500\, \rm g\, cm^{-1} s^{-1} \left(\frac{T_{\rm e}}{10^8 K} \right)^{5/2} \left(\frac{ln \Lambda}{40}\right)^{-1} \,
\end{equation}
where $\rm ln \Lambda$ is the Coulomb logarithm \citep{Sarazin1988}. 
We estimate $\epsilon$ based on our measured VSF on small scales, which is slightly different from $\epsilon$ estimated using velocities at the driving scale because the slopes of the VSFs are steeper than Kolmogorov. For gas properties, we use $T_{\rm e}=3$ keV and $n_{\rm e}=0.02 \rm cm^{-3}$ for Perseus \citep{Churazov2004}; for Abell 2597, we use $T_{\rm e}=2.7$ keV and $n_{\rm e}=0.06 \rm cm^{-3}$ \citep{Tremblay2012}; for Virgo, we use $T_{\rm e}=1.6$ keV and $n_{\rm e}=0.1 \rm cm^{-3}$ \citep{Zhuravleva2014}. 

According to direct numerical simulations \citep{Ishihara2016}, the gas viscosity affects pure hydrodynamic turbulence on scales that are larger than the Kolmogorov microscale (dashed grey line in Figure~\ref{fig:Irina}). Our detection of turbulence near and below the Kolmogorov microscale suggests that isotropic viscosity is suppressed in the ICM. 

For comparison, we also plot in Figure~\ref{fig:Irina} the measurement for Perseus using the X-ray surface brightness analysis, which assumes that density fluctuations follow the velocity field. Using the optical data, we are able to probe scales more than an order of magnitude smaller than X-ray observations of the same cluster. In fact, the electron mean free paths in the centers of Perseus and Virgo are $\sim 80$ pc and $\sim 8$ pc, respectively, about $1/3-1/2$ the size of our resolution in the two clusters. 

Figure~\ref{fig:Irina} also includes the best X-ray constraint on viscosity obtained from deep Chandra observations of the Coma cluster \citep{Zhuravleva2019}, where the mean free paths and the Kolmogorov microscales are larger. Our analysis based on the optical data probes the velocity field directly, and shows remarkable agreement with the conclusion of the X-ray surface brightness analysis. Both measurements support suppressed effective viscosity in the bulk intergalactic plasma, suggesting that the microphysics of the ICM is driven by magnetic fields operating below the Coulomb mean free path.

\subsection{Turbulence as a Heating Source}
It has been suggested that the dissipation of turbulence can balance radiative cooling in the centers of galaxy clusters based on the analysis of X-ray surface brightness fluctuations \citep[e.g.,][]{Zhuravleva2014}. The turbulent heating rate can be estimated as $Q_L\sim \rho v^3_L/L$ with L being the driving scale. Since our measured VSFs are in excellent agreement with the X-ray analysis within the scales that the X-ray observations probe (near the driving scale), the heating rate is similar when estimated using turbulence measured at the driving scale. 

However, as discussed previously, the slopes of the VSFs studied here tend to be steeper than Kolmogorov turbulence on small scales. If the steepening is caused by suppression of power on small scales, e.g., suppression of the nonlinear decay of gravity waves \citep{Bambic2018}, or AGN-driven turbulence being nonuniform, the actual dissipation rate should be somewhat lower than $Q_L$. On the other hand, if the steepening is a result of partial dissipation, the heating rate does not change. 

Another concern with AGN-driven turbulence as the main heating mechanism is that it may not propagate far enough to heat up the whole core \citep{Bambic2018b}. However, our VSFs reveal drivers at $\sim 20$ kpc in Perseus and Abell 2597, which we interpret as mainly reflecting the motions of the drivers themselves, not the propagations of turbulence from the very center of the cluster. Our analysis shows that turbulence at larger distances from the cluster centers can be generated ``in-situ'' by rising bubbles and possibly shocks as a result of AGN feedback. Therefore, our result is overall consistent with turbulence as an important heating mechanism.

\section{Conclusions and Final Remarks}\label{sec:conclusions}
Our study demonstrates the power of high resolution IFU observations in helping us understand the kinematics of multiphase gas. We show that AGN feedback is the main driver of turbulence in the centers of galaxy clusters. The result naturally serves as a test for numerical models of AGN feedback. In addition, it also serves as an excellent test for models of cool gas. Our detection of turbulence near the mean free path of the ICM supports suppressed effective viscosity. The slope of the VSF on small scales deviates from the classical Kolmogorov expectation, and points out directions for future theoretical and observational investigations. 

\section*{Acknowledgments}
We would like to thank Paul Duffell, Peng Oh, Christopher Reynolds, Anna McLeod and Andrea Antoni for helpful discussions. This work was partly performed at the Aspen Center for Physics, which is supported by National Science Foundation grant PHY-1607611. We acknowledge the technical support from the Scientific Computing Core of the Simons Foundation.

\appendix

\section{X-ray Analysis of Fluctuations in the Hot Gas in Perseus}

\begin{figure}
\centering
\includegraphics[width=0.45\linewidth]{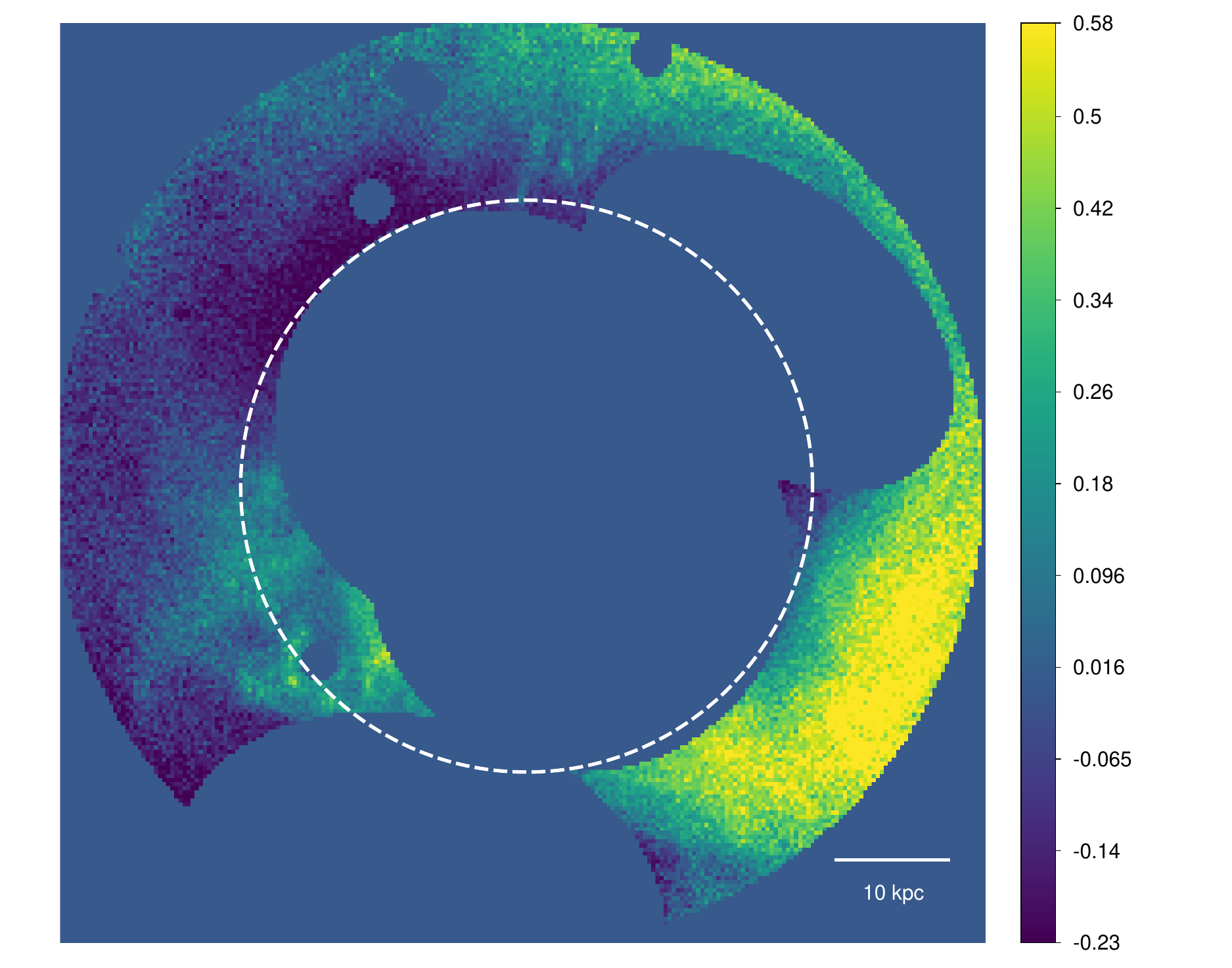}\\
\caption{Residual X-ray image of the central $r<40$ kpc region in the Perseus cluster used for the surface brightness fluctuation analysis shown in Figure~\ref{fig:Hitomi}. The central $r\lesssim25$ kpc region is excised from the analysis because it is dominated by bubbles and shocks produced by the AGN feedback.}\label{fig:PerseusX}
\end{figure}

We use deep Chandra observations of the Perseus cluster available in the archive. The initial data processing was done following the standard procedure \citep{Vikhlinin2005} that includes the filtering of high background periods, calculating the background intensity in each observation and application of the latest calibration corrections. The point sources are detected using the wvdecomp tool, and their significance are verified \citep{Zhuravleva2015}. These point sources are excised from the image accounting for the Chandra PSF. The residual image of the cluster (the image of fluctuations) is obtained from the initial cluster image divided by the best-fitting model of the mean surface-brightness profile. We calculate the power spectrum of the X-ray surface brightness fluctuations using the modified $\Delta$-variance method, which is suitable for non-periodic images with gaps \citep{Arevalo2012}. We re-project the spectra, correct them for the PSF and the unresolved point sources \citep{Churazov2012}. For Perseus, we analyzed the images in the 0.5-3.5 keV band. In this band, the resulting spectrum of fluctuations gives the 3D power spectrum of density fluctuations. Using a statistical linear relation between the power spectrum of density fluctuations and velocity \citep{Zhuravleva2014, Gaspari2014}, we obtained the power spectrum of gas motions in Perseus. 

The innermost $r\lesssim 25$ kpc region is dominated by the prominent structures associated with the bubbles of relativistic plasma and shocks around them \citep{Zhuravleva2015}. Therefore, we carefully select the region where the dynamics of the hot X-ray gas is probed. This region is shown in Figure~\ref{fig:PerseusX}. We effectively use fluctuations in the annulus $\sim25 - 40$ kpc. We additionally check the nature of fluctuations in this region \citep{Arevalo2016, Zhuravleva2016, Churazov2016} and confirm that most fluctuations in these regions are of isobaric nature.

%\clearpage
%\bibliography{library}
%\bibliographystyle{aasjournal}

\end{document}